\documentclass[
 reprint,
 amsmath,amssymb,
 aps,
prb,
floatfix,
]{revtex4-2}
\usepackage[percent]{overpic}
\usepackage{graphicx}
\usepackage{dcolumn}
\usepackage{bm}
\usepackage{hyperref}
\usepackage[mathlines]{lineno}
\usepackage{xcolor}
\usepackage[normalem]{ulem}
\usepackage[dvipsnames]{xcolor}

\begin{document}

\preprint{APS/123-QED}

\title{Reservoir Computing with a single Josephson junction\\}
\author{George Baxevanis}
\affiliation{School of Electrical and Computer Engineering, Aristotle University of Thessaloniki, 54124 Thessaloniki, Greece}

\author{Kathy L\"udge}
\affiliation{{Institut für Physik, Technische Universität Ilmenau, Weimarer Str. 25, 98693 Ilmenau, Germany}}

\author{Johanne Hizanidis}
\email[Correspondence email address:]{hizanidis@physics.uoc.gr}
\affiliation{Institute of Electronic Structure and Laser,
Foundation for Research and Technology-Hellas,
70013 Heraklion, Greece}
\affiliation{Institute of Nanoscience and Nanotechnology, National Center for Scientific Research, ``Demokritos'', 15341 Athens, Greece}

\date{\today}

\begin{abstract}
Physical reservoir computing exploits the nonlinear dynamics of a physical system to perform information processing tasks. Josephson junctions (JJs), as nonlinear superconducting devices with rich dynamical behavior, represent promising yet relatively unexplored candidates for reservoir computing.
In this work, we demonstrate for the first time that a single Josephson junction can be employed as a reservoir computing substrate without the use of an explicit delay loop. Using numerical simulations, we analyze the reservoir performance in different dynamical regimes and show that optimal performance is achieved when the JJ operates in a stable yet responsive regime.
Despite the absence of delayed feedback, the JJ exhibits sufficient memory through its intrinsic dynamics to achieve good performance on a chaotic time series prediction task. The underlying mechanism is analogous, at the dynamical level, to that of a driven nonlinear pendulum, highlighting the generality of the approach to other nonlinear oscillators. In addition, we explore an alternative input masking approach based on continuous modulation, highlighting its compatibility with practical implementations.
These results establish Josephson junctions as a viable and efficient platform for reservoir computing and open the way to ultrafast, low-dissipation hardware realizations.

\end{abstract}

\maketitle

\section{\label{sec:Intro} Introduction }

Reservoir computing provides a framework for temporal information processing by exploiting the dynamics of nonlinear systems while requiring training only at the level of a linear readout~\cite{cucchi2022hands,yan2024emerging}. The approach originates from the echo state network~\cite{jaeger2001echo} and liquid state machine~\cite{maass2002real} paradigms, where a fixed dynamical system, the reservoir, transforms input signals into a high-dimensional representation. The resulting states encode both current and past inputs, enabling efficient learning of temporal tasks through simple output weight optimization~\cite{lukovsevivcius2009reservoir}.

A key feature of reservoir computing is the nonlinear transformation performed by the reservoir, which maps input signals into a high-dimensional state space while retaining information about their temporal history. This combination of nonlinearity and memory enables efficient learning through a simple linear readout. Owing to these properties, reservoir computing has been widely applied to the prediction and analysis of chaotic and spatiotemporally complex dynamical systems~\cite{PhysRevLett.120.024102,neofotistos2019machine, Jaurige2022}.

These ideas have led to the paradigm of physical reservoir computing~\cite{tanaka2019recent,stepney2024physical,nakajima2021reservoir}, in which the reservoir is realized by a physical dynamical system. In this setting, the number of tunable parameters is typically reduced, as the system’s behavior is largely determined by its intrinsic physical properties rather than freely adjustable network weights. As a result, the required computational features emerge directly from the underlying dynamics, providing a natural and hardware-efficient implementation.

A wide range of physical systems has been explored as substrates for reservoir computing. Memristor-based reservoir computing exploits the intrinsic short-term resistance dynamics of volatile memristive devices to perform temporal information processing directly in hardware, enabling compact and scalable implementations for tasks such as pattern recognition and time series prediction~\cite{du2017reservoir,jang2024review}. Photonic reservoir computing~\cite{ABD26}, in turn, has attracted significant interest due to its large bandwidth and inherent parallelism with both integrated and optoelectronic systems demonstrating efficient temporal signal processing~\cite{cai2020modulation,skontranis2022time,GOL25}. These photonic systems are usually delay-based setups \cite{appeltant2011information,hulser2022role} and can be efficiently tuned to adapt to different tasks and timescale requirements, as long as the bifurcation structure of the underlying dynamical system are known\cite{muhlnickel2024influence,dong2025time}. 
Reservoir computing has furthermore been implemented in a variety of unconventional physical substrates, including DNA-based and biology-inspired systems~\cite{goudarzi2013dna,flynn2023seeing}, electrochemical systems~\cite{shingu2023electrochemical}, and electromechanical platforms such as piezoelectric resonators~\cite{wang2026piezoelectric} or magnonic systems~\cite{XIO26}.   The choice of a specific reservoir depends on the envisioned application, as the reservoir’s characteristic timescales and signal domain should be compatible with those required by the target task. In this respect, our work explores single Josephson Junctions as a reservoir computing platform, with the aim of understanding system-specific properties and identifying suitable modes of operation, rather than pursuing record performance (see also a recent perspective on in-materio physical reservoirs \cite{SIN26}). 

Superconducting and quantum platforms are particularly compelling due to their ultrafast dynamics, low dissipation, and inherent nonlinearity, and the high dimensionality of quantum systems has been already exploited for temporal information processing and chaotic time-series prediction~\cite{gyurik2025quantum,kent2026superconducting,Steinegger2025,CIN24}. 


At the core of many such superconducting architectures lies the fundamental nonlinear element of superconducting electronics: the Josephson junction. JJs have attracted sustained interest due to their ability to exhibit extremely fast voltage dynamics and rich nonlinear behavior. Since their discovery~\cite{JOS62,LIK1986}, JJs have become key building blocks in a wide range of modern technologies, including quantum science, superconducting electronics, and high-frequency applications. Using those elements as a reservoir computer would allow to design multi purpose applications using a mature and well spread technology.  

Recently, Josephson junctions have been explored in the context of superconducting neuromorphic computing~\cite{schneider2022supermind,schneider2025self}. By nature, JJs exhibit excitable, neuron-like behavior, and when coupled, they can emulate more complex neuronal features such as action potentials, bursting, and first-spike latency~\cite{crotty2010josephson,chalkiadakis2022dynamical,baxevanis2025inductively}.
While these studies largely adopt a biomimetic perspective, treating JJs as physical neurons that reproduce specific biological functionalities, an alternative approach is to exploit their dynamics more generally for computation, as in reservoir computing. 

To date, JJs have primarily been employed in reservoir computing architectures based on Josephson transmission lines (JTLs), where sequentially biased junctions collectively form the reservoir~\cite{rowlands2021reservoir,watanabe2024numerical}. These systems have been evaluated on tasks such as parity checking, channel equalization, and image recognition.
In contrast, here we investigate the computational capabilities of a reservoir consisting of a single Josephson junction. This minimal configuration allows us to directly relate dynamical properties to computational performance. Furthermore, even when accounting for cryogenic operation at $4.2\,\mathrm{K}$, JJs retain a significant energy advantage~\cite{holmes2013energy}, reinforcing their potential as efficient neuromorphic computing elements. 


The manuscript is organized as follows: We first introduce the Josephson junction model and its dynamical properties, followed by the reservoir computing framework and implementation. We then present a systematic analysis of the reservoir performance in different dynamical regimes, including hyperparameter studies and the system’s response in time and phase space. An alternative input masking approach based on sinusoidal modulation is also investigated. Finally, we summarize the main findings in the Conclusions and outline possible directions for future work.

\section{\label{sec:Model} The Josephson Junction Reservoir}
This section is divided into two parts. First, we present the Josephson junction model and the governing equations describing its dynamics. We then discuss how the JJ can be implemented as a time-multiplexed reservoir, where the choice of operating regimes is directly informed by its dynamical behavior and underlying bifurcation structure. In this sense, the intrinsic dynamics of the JJ play a central role in shaping and optimizing the reservoir operation.
\subsection{Josephson junction dynamics}
A Josephson junction is made up of two superconducting elements with a separating nonsuperconducting gap material between them (e.g. an insulator or a normal metal). Each superconductor can be described by a quantum mechanical wave function \(\psi_1 e^{i\phi_1}\) and \(\psi_2 e^{i\phi_2}\) accordingly. 
When a dc current is applied to the JJ, which is less than a critical current threshold \(I_c\), no voltage develops across it. In circuit terminology, this means that the JJ acts as if it has zero resistance. Despite the absence of a voltage drop, a constant phase difference $\phi = \phi_1 - \phi_2$ develops between the two superconducting elements, with $\phi$ satisfying the Josephson current-phase relation
$I = I_c \sin\phi$.
Once the input current exceeds the critical threshold $I_c$, a voltage is generated across the Josephson junction, according to 
$V = {\hbar}/{(2e)}{d\phi}/{dt} = {\Phi_0}/{(2\pi)} {d\phi}/{dt}$, 
where $\Phi_0$ is the flux quantum, $t$ denotes the time, $e$ is the electron charge and $\hbar$ is the Planck’s constant.

Josephson junctions play a fundamental role in enabling and controlling macroscopic quantum phenomena, as further highlighted by recent Nobel Prize-winning advances in superconducting circuits~\cite{schirber2025nobel}. Notable examples include superconducting quantum interference devices (SQUIDs), which enable the detection of extremely small magnetic fields~\cite{hizanidis2018flux}, superconducting quantum circuits, where JJs serve as the building blocks of qubits~\cite{martinis2004superconducting}, and superconducting metamaterials, in which they are used to tune electromagnetic properties~\cite{cai2024effects}.

The Josephson junction can be described within the framework of the resistively and capacitively shunted junction (RCSJ) model~\cite{LIK1986}. In this approach, the junction is represented by an equivalent parallel circuit, where, in addition to the supercurrent $I$, a capacitor $C$ 
accounts for the displacement current and a resistor $R$ accounts for an effective current associated with quasiparticle contributions~\cite{de2018superconductivity,pomorski2016justification}. Applying Kirchhoff's current and
voltage laws yields the following equation:
\begin{equation}
 \frac{C \hbar}{2e} \ddot{\phi}+ \frac{\hbar}{2eR} \dot{\phi} + I_c \sin\phi  = I_B,
\label{eq:RCSJ_eq_2}
\end{equation}
where $I_B$ is the bias current applied to the device.

To provide a sense of the physical scales and operating characteristics of a Josephson junction, typical device dimensions are on the order of $1\,\mu\mathrm{m}$, with critical currents $I_c$ ranging from $1\,\mu\mathrm{A}$ to $1\,\mathrm{mA}$ and characteristic voltages $I_cR\approx1\,\mathrm{mV}$~\cite{STR94}. Taking into account the Josephson constant, $2e/h = 4.83 \times 10^{14}\,\mathrm{Hz/V}$, such voltage scales correspond to operating frequencies on the
order of $10^{11}\,\mathrm{Hz}$, i.e., in the sub-THz to THz regime.

We use Eq.~\eqref{eq:RCSJ_eq_2} and nondimensionalize the system as follows
\begin{equation}
  \phi'' + \alpha \phi' + \sin\phi = I_\mathrm{dc},
\label{eq:RCSJ_eq_alpha}
\end{equation}
where parameter $\alpha=\sqrt{\hbar/(2e I_c R^2 C)}$ is the dimensionless damping, $I_\mathrm{dc}=I_B/I_c$ is the dimensionless applied dc current and derivatives are now with respect to $\tau=\sqrt{2 e I_c / (\hbar C)}\, t$. For the analysis, we choose $\alpha>0$ for physical plausibility and $I_\mathrm{dc} \geq 0$ (using $I_\mathrm{dc} \leq 0$ would lead to similar results by changing $\phi$ with $-\phi$). 
If we set the variable $\mathsf{V} = \phi'$ denoting the dimensionless voltage, we can write Eq.~\eqref{eq:RCSJ_eq_alpha} as a system of two first-order differential equations
\begin{align}
\phi' &= \mathsf{V} \label{eq:phidot} \\
\mathsf{V}'    &= I_\mathrm{dc} - \sin\phi - \alpha \mathsf{V}. \label{eq:Vdot}
\end{align}

\begin{figure}[tb]
\setlength{\unitlength}{1\linewidth}
\begin{picture}(1,0.50)
\includegraphics[width=0.85\columnwidth]{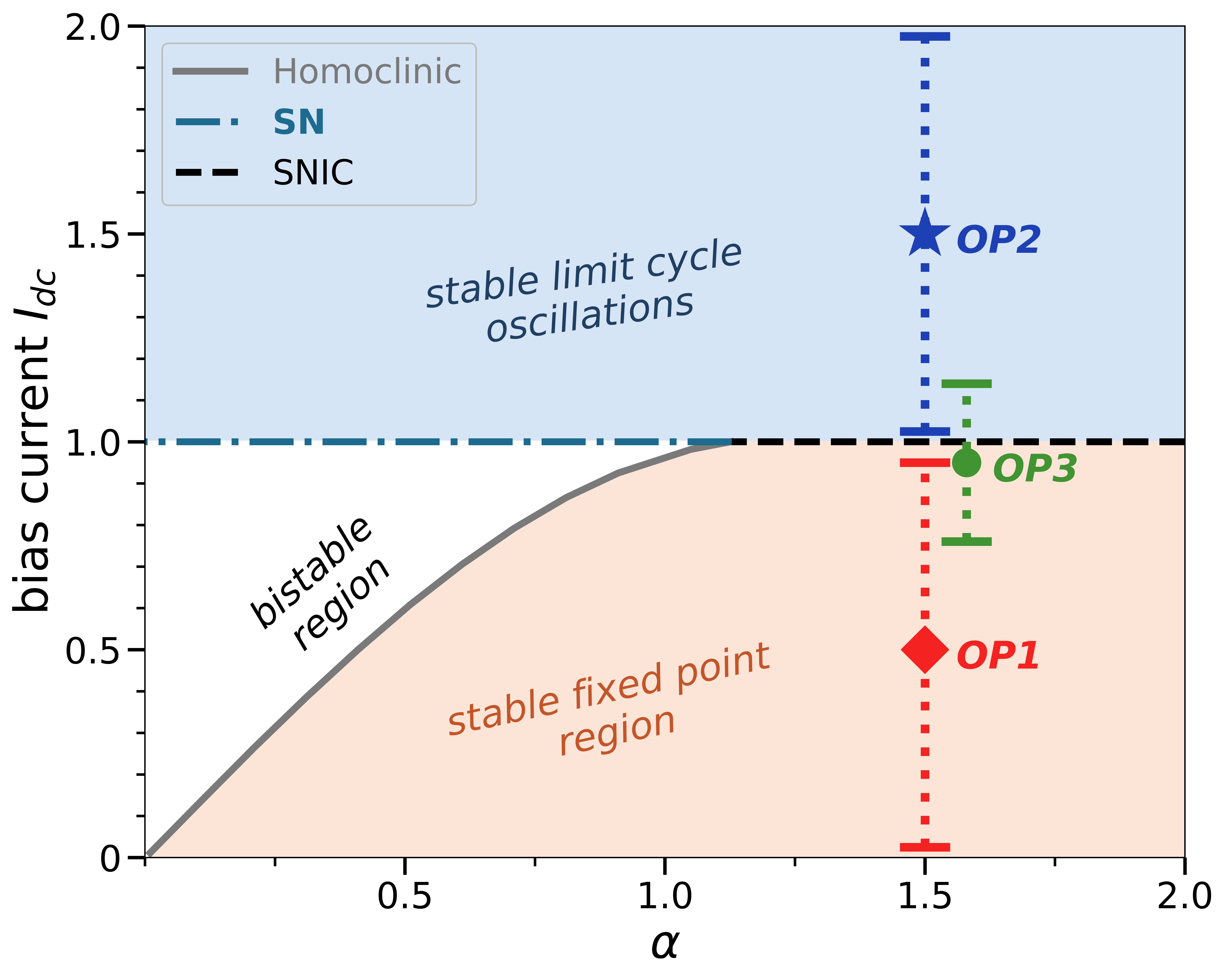}
  \label{fig:bif_diagram}
\end{picture}
\caption{Stability diagram of the Josephson junction in the $(\alpha, I_\mathrm{dc})$ plane. The homoclinic, saddle-node (SN), and saddle-node bifurcation on invariant circle (SNIC) bifurcation curves define regions of stable fixed point (orange shading), bistability (white), and stable limit cycle oscillations (blue shading). The selected operation points (OP1–OP3) used in the Sec.~\ref{sec:results} are marked as red diamond, blue star and green circle together with range of total injected current used after hyperparameter optimization. (OP3 has been shifted for clarity.)}
\label{fig:bif_diagram}
\end{figure}

Figure~\ref{fig:bif_diagram} shows the well-known stability diagram for the JJ in the $(\alpha, I_\mathrm{dc})$ parameter space. Three distinct bifurcations are present: a saddle-node (SN), a homoclinic, and a saddle-node on an invariant cycle (SNIC) bifurcation. These define three dynamical regimes, namely a stable fixed point, a stable limit cycle, and a bistable region where both coexist. In the following subsection, we demonstrate how this diagram guides the selection of parameters for the reservoir.

\begin{figure}[tb]
\includegraphics[width=\columnwidth]{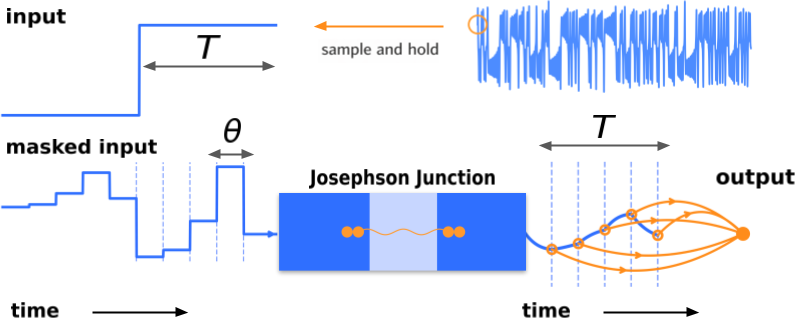} 
\caption{\label{fig:jj_rc} Time-multiplexed JJ reservoir scheme. The Lorenz 63 $X$ time series input is held constant over a clock cycle $T$, multiplied by a random mask to generate $N$ virtual nodes (marked by the vertical dashed lines) with duration $\theta$ each, and injected into the Josephson junction. The resulting dynamical response is sampled to form the reservoir states (marked by the full circles), which are linearly combined to produce the output.}
\end{figure}
A schematic of the JJ reservoir is shown in Fig.~\ref{fig:jj_rc} and a detailed description is provided in the following subsection.

\subsection{Time-multiplexed reservoir computing}
Before delving into the JJ reservoir implementation, we briefly review the fundamentals of time-multiplexed reservoir computing, the standard framework for physical realizations of reservoir computing (RC) which is also visualized in Fig.~\ref{fig:jj_rc}.

Time-multiplexed reservoir computing provides an efficient implementation of reservoir computing by replacing the spatially extended network
of the conventional RC paradigm with a single nonlinear dynamical system \cite{appeltant2011information}. Instead of using multiple physical nodes, a single node is driven sequentially in time, thereby generating a set of \emph{virtual nodes} through temporal sampling. It is noted that delay-based schemes are widely used for time multiplexed RC setups, introducing, thus, an additional timescale and enhancing the memory of the system~\cite{hulser2022role,muhlnickel2024influence}. Our JJ reservoir implementation has no internal delay and short term memory is provided only by the dynamical system response. While it is possible to externally add memory via post- or preprocessing \cite{JAU25,jaurigue2024reducing}, we study the solitary JJ and explore its RC abilities.   

In this framework, the discrete input signal $\mathrm{I}(k)$, at time $k$,
is held constant over a time window $T$ (clock cycle) and multiplied by a predefined random mask sequence $m(j)$, with $j = 1, \dots, N$, where $N$ denotes the number of virtual nodes. This produces a time-dependent input signal
\begin{equation}
    J(j,k) = \mathrm{I}(k)\, m(j), \label{eq:I}
\end{equation}
which is injected sequentially into the reservoir (see Appendix \ref{appendix_a} for details of the task). The random mask takes values in the interval $(-1,1)$, drawn from a uniform probability distribution. There are several ways in which the masked input can enter the Josephson junction. A natural choice is to inject it through the bias current $I_\mathrm{dc}$, as this is readily accessible in experiments. Moreover, since $I_\mathrm{dc}$ serves as the second control parameter in the stability diagram, it provides a convenient starting point and offers intuitive insight into the system’s expected behavior.

A straightforward implementation consists of adding the masked input signal, scaled by a factor $K_{\text{inj}}$, such that Eq.~\eqref{eq:Vdot} becomes
\begin{equation}
   \mathsf{V}'= I_\mathrm{dc} + K_{\text{inj}}\, J-\sin\phi - \alpha \mathsf{V}
. \label{eq:JJ_system_Kinj1}
\end{equation}

Each masked input value is applied for a duration 
\begin{equation}
    \theta = \frac{T}{N},
\end{equation}
which defines the temporal spacing between consecutive virtual nodes. The continuous time response of the dynamical system is then sampled at intervals of $\theta$, yielding a sequence of reservoir states
\begin{equation}
    x_j(k) = x\big(t = kT + j\theta\big), \quad j = 1,\dots,N.
\end{equation}

Collecting these responses over $K$ input steps forms the state matrix
\begin{equation}
    \mathbf{S} \in \mathbb{R}^{K \times (N+1)},
\end{equation}
where an additional bias term is typically included. The output is obtained via a linear readout
\begin{equation}
    \hat{\mathbf{y}} = \mathbf{S}\mathbf{w},
\end{equation}
where the output weights $\mathbf{w}$ are determined by ridge regression
\begin{equation}
    \mathbf{w} = \left(\mathbf{S}^\top \mathbf{S} + \lambda \mathbf{I}\right)^{-1} \mathbf{S}^\top \mathbf{y}.
\end{equation}
Here, $\lambda$ is the regularization parameter used to prevent overfitting.
The reservoir performance is evaluated using the normalized root mean squared error (NRMSE), defined as
\begin{equation}
\mathrm{NRMSE} =
\frac{
\sqrt{\frac{1}{N_s}\sum\limits_{k=1}^{N_s}(\mathrm{y}[k]-\hat{\mathrm{y}}[k])^2}
}{
\sqrt{\mathrm{var}(\mathbf{y})}
},
\label{eq:NRMSE}
\end{equation}
where $\mathrm{y[k]}$ denotes the target signal, $\hat{\mathrm{y}}[k]$ the reservoir prediction, $N_s$ the number of training or testing samples, and $\mathrm{var}(\mathbf{y})$ the variance of the target signal. For our simulations, we always average over multiple mask realizations to reduce the impact of a single mask and therefore always discuss mean NRMSE values in the following.

The performance of the time multiplexed RC system depends on several key hyperparameters. The number of virtual nodes $N$ and the node spacing $\theta$ determine the effective dimensionality and temporal resolution of the reservoir. The injection strength $K_{\text{inj}}$ controls how strongly the input signal perturbs the intrinsic JJ dynamics, thereby influencing the explored region of the stability diagram. The mask $m(j)$ (and thus the masked input $J$) enriches the input representation by introducing temporal diversity. Finally, the regularization parameter $\lambda$ balances the trade-off between fitting accuracy and generalization in the readout layer.

In the following sections, we investigate the injection scheme given by Eq.~\eqref{eq:JJ_system_Kinj1} in different regions of the stability diagram of the JJ dynamics in order to search of optimal parameters for the JJ reservoir. As a benchmark task, we use one-step-ahead prediction of the Lorenz 63 system~\cite{Lorenz2004}, where the objective is to predict the $X$ variable according to $X(t) \to X(t+\Delta t)$~\cite{jaurigue2024reducing}. Details of the Lorenz system, numerical integration, and data preparation are provided in Appendix~\ref{appendix_a}.
This input is scaled with the injection strength $K_{\mathrm{inj}}$, which allows us to control the operating regime of the JJ reservoir within the desired regions of the stability diagram. We refer to a specific choice of parameters corresponding to a given dynamical regime as an \emph{operating point} (OP).
The selection of the latter is guided by the stability diagram in Fig.~\ref{fig:bif_diagram} where OPs are marked with colored symbols. 

We first identify an appropriate value for the damping parameter. In principle, any value of $\alpha$ could be chosen, however, we avoid the regime $\alpha < 1.2$, as it involves two bifurcation mechanisms and supports three distinct dynamical states, which may introduce unnecessary complexity into the reservoir dynamics.
In particular, the bistable regime may lead to different reservoir responses under similar input conditions, depending on the dynamical state of the system, potentially compromising the consistency of the reservoir response.
Therefore, we restrict our attention to $\alpha > 1.2$, and in particular choose $\alpha=1.5$, which allows the JJ to access two dynamical regimes: a stable fixed point and a stable limit cycle. 

For the bias current, we consider three representative values corresponding to distinct dynamical regimes in Fig.~\ref{fig:bif_diagram}. Specifically, OP1 is marked with a red diamond and lies within the stable fixed point regime, OP2 is marked with a blue star and lies within the oscillating regime, and OP3 is marked with a green circle and lies just beneath the SNIC bifurcation at $I_\mathrm{dc} = 1$. The inclusion of OP3 close to the SNIC bifurcation is motivated by the reported enhanced computational performance of reservoir computing near the edge of stability~\cite{carroll2020reservoir}. The vertical dashed lines mark the range of the driving current during reservoir computing operation for each operation point.

\section{Reservoir Dynamics and Performance} \label{sec:results}
In this section, we analyze the performance of the JJ reservoir across different dynamical regimes. We begin with hyperparameter scans to identify suitable operating conditions, followed by an examination of the reservoir response in both time series and phase space representations. This analysis reveals how the intrinsic JJ dynamics affects the computational capacity of the reservoir. Finally, we investigate an alternative to the random masking scheme and its impact on performance.

\subsection{\label{sec:hyperparam} Hyperparameter Tuning}
\begin{figure}[tb]
\centering
\vspace{1em}
\begin{overpic}[width=0.8\columnwidth]{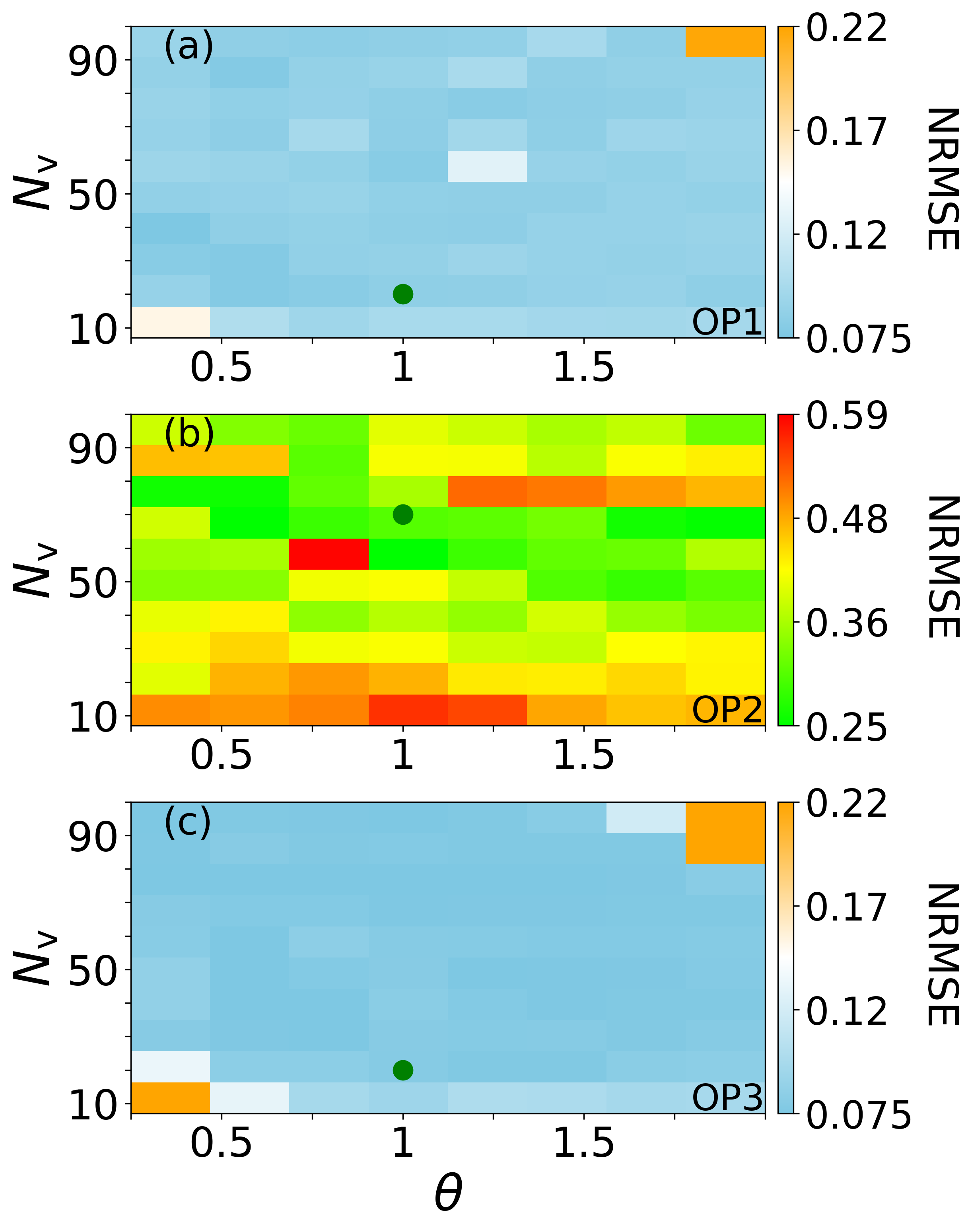}
\end{overpic}
\caption{Prediction performance $\text{NRMSE}_\text{test}$ for Lorenz one step ahead prediction as a function of virtual node number $N_V$ and node processing time $\theta$. The plotted values are
averaged over five realizations with different random masks. (a) OP1 with $I_\mathrm{dc}=0.5$, $K_{\mathrm{inj}}=0.25$ and $\lambda=10^{-13}$, (b) OP2 with $I_\mathrm{dc}=1.5$, $K_{\mathrm{inj}}=0.25$ and $\lambda=10^{-13}$, and (c) OP3 with $I_\mathrm{dc}=0.95$, $K_{\mathrm{inj}}=0.1$ and $\lambda=10^{-13}$. 
The green dots indicate the best performance values found for parameter combinations: $N_V=20$ for (a), $N_V=70$ for (b), and $N_V=20$ for (c), with $\theta=1$ in all cases.}
\label{fig:Nodes_theta_scan_col}
\end{figure}

The stability diagram presented above will guide the selection of representative dynamical regimes, while the parameter scans in Fig.~\ref{fig:Nodes_theta_scan_col} identify suitable values of $N_V$ and $\theta$ for each operating point.
We begin our analysis for OP1 (see red diamond in Fig.\ref{fig:bif_diagram}), where we set the dc current to $I_\mathrm{dc}=0.5$ and the injection strength to $K_{\mathrm{inj}}=0.25$,
to ensure that, under the applied input, the JJ remains in the stable fixed point regime (orange shading). 

The virtual node spacing, $\theta$, is known to be related to the intrinsic timescales of the reservoir \cite{dong2025time}, while the number of virtual nodes, $N_V$, determines the effective dimensionality of the network. For the JJ considered here, the intrinsic dynamics exhibit characteristic timescales ranging from a few time units for the local asymptotic relaxation to approximately $10$ time units for the full transient response. The optimal value $\theta=1$ is therefore comparable to the local relaxation timescale, while the corresponding clock cycles $T=N_V\theta$ can extend beyond the characteristic transient timescale.

To identify suitable values of $N_V$ and $\theta$, we evaluate the prediction performance in
this parameter space, as shown in Fig.~\ref{fig:Nodes_theta_scan_col}. The product of $N_V$ and $\theta$ determines the length of the clock cycle $T$, so combinations with the same product lie on constant-$T$ hyperbolas in the $(\theta, N_V)$ parameter space. For OP1 (red diamond in Fig.\ref{fig:bif_diagram}), we observe that a relatively small number of virtual nodes, $N_V = 20$, is sufficient, and we select $\theta = 1$, for which a good value of $\text{NRMSE}_\text{test} = 0.085\pm \textrm{5.2e-06}$  is achieved (marked by the green dot in Fig.~\ref{fig:Nodes_theta_scan_col}a).

The second operating point OP2 (blue star in Fig.\ref{fig:bif_diagram}) is chosen such that the JJ reservoir remains within the oscillating regime and the results are presented in Fig.~\ref{fig:Nodes_theta_scan_col}b. We choose a bias current of $I_\mathrm{dc} = 1.5$, thereby avoiding crossing the infinite-period (SNIC) bifurcation during operation. In this regime, the reservoir prediction performance 
is notably degraded, as can be seen by the high NRMSE values (yellow-red regions in Fig.~\ref{fig:Nodes_theta_scan_col}b). The best performance for OP2 is obtained for $N_V = 70$ and $\theta = 1$, yielding a mean $\mathrm{NRMSE} = 0.25$.

For the third operating point, OP3, (green circle in Fig.\ref{fig:bif_diagram}) the results are shown in Fig.~\ref{fig:Nodes_theta_scan_col}c. Here we allow the JJ to transition between the stable limit cycle and stable fixed point regimes (see green dot in Fig.~\ref{fig:bif_diagram}). The underlying idea is that crossing the SNIC bifurcation line enriches the reservoir dynamics and enhances its capacity for nonlinear transformations, while maintaining a consistent response to the input. We therefore choose a dc bias current of $I = 0.95$, such that the mean total current lies within the stable fixed point regime.
As can be seen in Fig.\ref{fig:Nodes_theta_scan_col}c, for $N_V=20$ and $\theta=1$ the performance is $\text{NRMSE}_\text{test}=0.077\pm \textrm{1e-05}$ . 

The choices of the regularization parameter $\lambda$ and the injection strength $K_{\mathrm{inj}}$, initially fixed based on the considerations above, 
are subsequently validated through parameter scans presented in Fig.~\ref{fig:Lambda_Kinj_scan} in Appendix \ref{appendix_b}. After the final choice of all parameters, the resulting performance over 15 realizations is $\text{NRMSE}_\text{test}=0.087  \pm 0.0047$ for OP1, $\text{NRMSE}_\text{test}=0.28  \pm 0.029$ for OP2 and $\text{NRMSE}_\text{test}=0.073 \pm  0.009$ for OP3. Although OP3 achieves a better mean performance, we favor OP1 because it operates in a more robust parameter regime, as discussed in Secs. ~\ref{sec:JJ_Response} and ~\ref{sec:Bias_Input_Effect}. To further elucidate the effect of the parameters on the performance, we examine the dynamical response of the JJ to the driving input signal in the following subsection.

\subsection{\label{sec:JJ_Response} JJ Reservoir Response}
\begin{figure*}[t]
\begin{overpic}[width=1\textwidth]{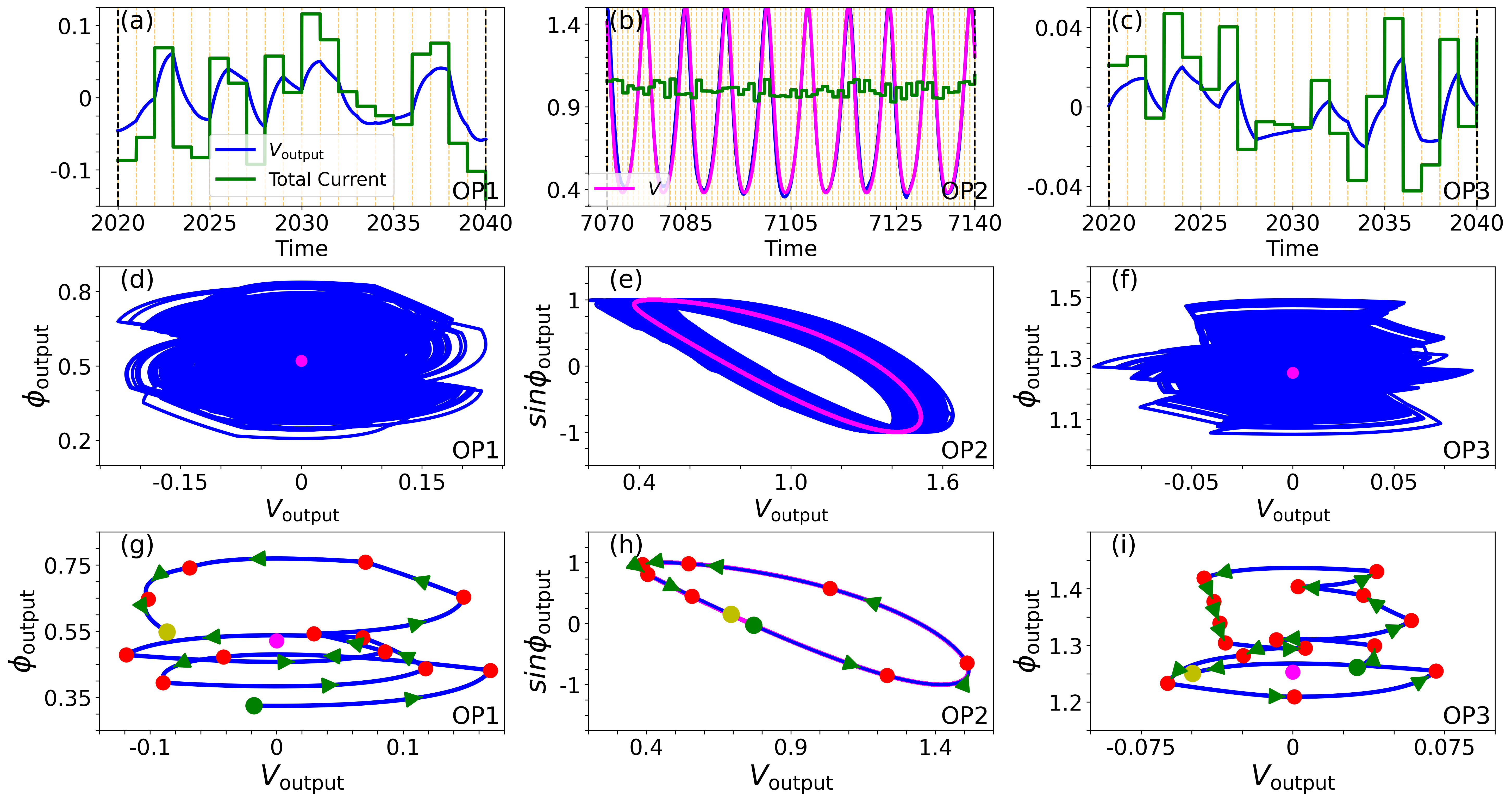}
\end{overpic}
\caption{Top row: Depiction of the $V_{\mathrm{output}}$ (blue) time series for the three operation points with the corresponding virtual nodes (orange), together with the total input current (green), from which the mean over the displayed window has been subtracted for clarity (Figs.~(a)--(c)). Middle row: Phase-space trajectories (blue) in the $(V_{\mathrm{output}},\phi_{\mathrm{output}})$ plane, or in the $(V_{\mathrm{output}},\sin\phi_{\mathrm{output}})$ plane for OP2, over many clock cycles [Figs.~(d)--(f)]. Bottom row: Zoomed-in trajectories for part of a clock cycle [Figs.~(g)--(i)]. The green and yellow dots mark the start and end points of the trajectory, green arrows indicate the direction of evolution, and red dots show when a new input is injected into the JJ reservoir. For OP2, the unperturbed limit cycle is shown in magenta. In Figs.~(d), (f), (g), and (i), the magenta dots denote the stable fixed points corresponding to the mean total input current. Parameters for OP1 are $N_V=20$, $\theta=1$, $K_{\mathrm{inj}}=0.25$, $\lambda=10^{-13}$, and $I_\mathrm{dc}=0.5$; for OP2, $N_V=70$, $\theta=1$, $K_{\mathrm{inj}}=0.15$, $\lambda=10^{-13}$, and $I_\mathrm{dc}=1.5$; and for OP3, $N_V=20$, $\theta=1$, $K_{\mathrm{inj}}=0.1$, $\lambda=10^{-13}$, and $I_\mathrm{dc}=0.95$.}
\label{fig:V_timeseries_V_Phi_plot}
\end{figure*}

In order to understand the significantly poorer performance of OP2 compared to OP1 and OP3, Fig.~\ref{fig:V_timeseries_V_Phi_plot} shows voltage timeseries (panels a-c), phase-space projections (panels d-f), and zoomed-in phase space trajectories in the $(V_{\mathrm{output}} ,\phi_{\mathrm{output}})$ plane (panels g–i) for the three operating points, arranged column-wise.

From the time series, it is evident that the JJ response (blue) closely follows the total current (green) in OP1 and OP3. In contrast, for OP2, the dynamics is dominated by the limit cycle oscillations (the unperturbed limit cycle is shown in magenta), which are only weakly influenced by the input. The vertical dashed orange lines mark the positions of the virtual nodes.

From the corresponding phase-space representations in Fig.~\ref{fig:V_timeseries_V_Phi_plot}d–f, we observe that in OP1 and OP3 the dynamics explore a broader region around the fixed point (magenta dot). In contrast, for OP2 the system response remains confined close to the unperturbed limit cycle (magenta orbit). Consequently, the JJ reservoir in OP2 explores the state space less effectively, resulting in reduced variability of the sampled states and, ultimately, poorer performance.

To illustrate the short-term evolution of the reservoir and the effect of successive inputs, Fig.~\ref{fig:V_timeseries_V_Phi_plot}g-i shows example trajectories over a limited number of virtual nodes for the three operating points. Starting from the initial state (dark green dot), the system evolves along the direction indicated by the green arrows until it reaches the final state (yellow dot). The red dots mark the instances at which new input values are injected into the reservoir.
In OP1 and OP3, each input results in noticeable deviations in the trajectory, leading to a richer exploration of the phase space. In contrast, for OP2, the trajectory remains largely confined near the limit cycle (magenta), indicating that the input has only a weak effect on the system dynamics.

Among the three operating points considered, OP1 provides the most robust performance on the time-series prediction task. Although OP3, where the SNIC bifurcation is traversed, exhibits even better mean performance, its proximity to the region of more complex dynamics leads to larger standard deviations in the NRMSE (see Fig.~\ref{fig:Idc_vs_Kinj}b). This raises the question of whether crossing the infinite-period bifurcation gives rise to a qualitatively richer system response and relates to the broader discussion of whether reservoir computing performance is enhanced near the edge of stability~\cite{carroll2020reservoir}.

By comparing the direct response to the input for OP1 and OP3 in Fig.~\ref{fig:V_timeseries_V_Phi_plot}a,c, we observe strong similarities. Despite the bifurcation being crossed in OP3, the input perturbation is not sufficient to induce sustained limit cycle oscillations. In both cases, the JJ reservoir responds nonlinearly to the injected input current while remaining effectively in a fixed point-like regime.
Only if the system is allowed to evolve over longer times between successive input injections would additional oscillatory behavior emerge.

To put the performance of the JJ reservoir into perspective, we compare the best result obtained at our selected operating point, OP1, with other reservoir computing implementations for the same Lorenz prediction task (see Appendix \ref{appendix_a} for all the details concerning the task). A delay-based, time-multiplexed reservoir with $250$ virtual nodes and a static nonlinearity is reported to reach $\mathrm{NRMSE}_{\mathrm{test}} = 0.02$~\cite{jaurigue2024reducing} while with $20$ nodes a value of $0.09$ is achieved. The latter is comparable to our results, however we note that our implementation does not have a delay line. Simulations based on a time-multiplexed quantum-dot laser reservoir with feedback have reached performance values of $\mathrm{NRMSE}_{\mathrm{test}} = 0.014$ with $30$ nodes~\cite{dong2025time}. There, the effect of feedback and internal dynamics are combined which explains the better performance value. Further improvements have been reached using spin-VCSEL systems with optical injection and feedback ($\mathrm{NRMSE}_{\mathrm{test}} = 0.006$ with $N_V = 60$)~\cite{muhlnickel2024influence}.  Consequently, our JJ reservoir based on a solitary element is compatible with other physical implementations as far as the avaiable nonlinear transformations are concerned. It is limited in memory which could, however, be incorporated into the JJ reservoir through delayed input~\cite{PIC25,parlitz2024learning, jaurigue2024reducing, jaurigue2021reservoir} or delayed output~\cite{JAU25,jaurigue2024reducing},  leading to further performance improvements. Here, however, we deliberately focus on the capabilities of a single JJ reservoir without additional delay-based augmentation.

\subsection{\label{sec:Bias_Input_Effect}Effect of Bias Current and Input Scaling}
\begin{figure}[tb]
\centering
\begin{overpic}[width=0.8\columnwidth]{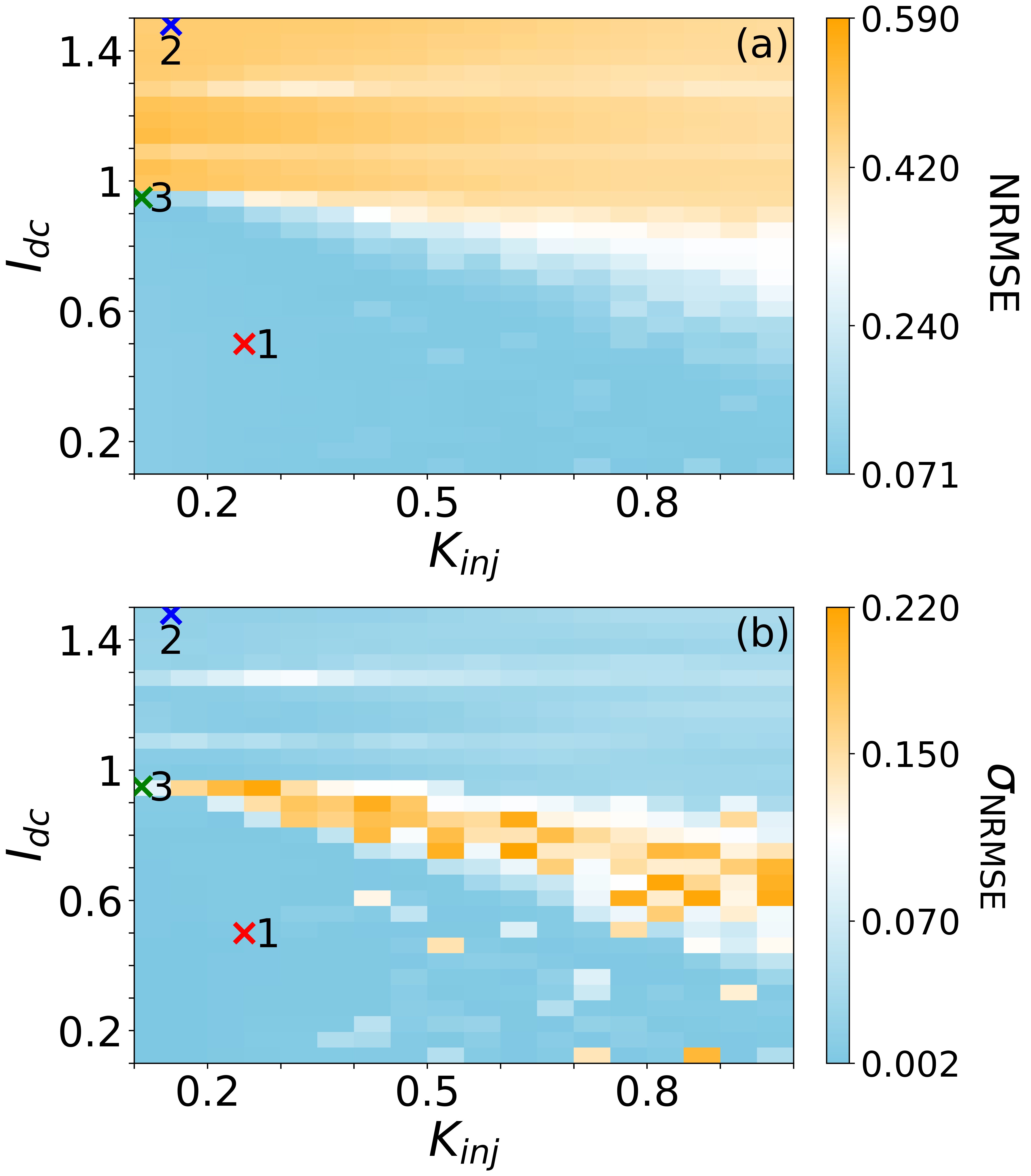}
\end{overpic}
\caption{Scan over different combinations of $I_\mathrm{dc}$ and $K_{\mathrm{inj}}$ based on Eq.~\eqref{eq:JJ_system_Kinj1}, computed over $15$ realizations, where the input is added to the dc current. Parameters are $N_V=20$, $\theta=1$, $\lambda=10^{-13}$, and damping $\alpha=1.5$. Panel (a) shows the mean $\mathrm{NRMSE}_\mathrm{test}$, while panel (b) shows the corresponding standard deviation. The crosses denote the operating points selected earlier (red: OP1, blue: OP2, green: OP3). Note that for OP2, the optimal value was $N_V=70$, rather than $N_V=20$ used in this scan.}
\label{fig:Idc_vs_Kinj}
\end{figure}

The dynamical insights obtained in the previous subsection suggest that, in OP2, the limit cycle oscillations are only weakly affected by the input. We therefore investigate whether increasing the injection strength or adjusting the dc bias current can improve the performance of the JJ reservoir.
Figure~\ref{fig:Idc_vs_Kinj} shows a parameter scan over $I_{\mathrm{dc}}$ and $K_{\mathrm{inj}}$, based on $20$ realizations. The mean $\mathrm{NRMSE}_{\mathrm{test}}$ is presented in Fig.~\ref{fig:Idc_vs_Kinj}a, while the corresponding standard deviation is shown in Fig.~\ref{fig:Idc_vs_Kinj}b. Since the previous analysis of operating points yielded satisfactory performance for $N_V = 20$, $\theta = 1$, and $\lambda = 10^{-13}$ (in OP1 and OP3), these values are kept fixed throughout this parameter scan.

The first observation is that, as the dc current approaches the SNIC bifurcation at $I_{\mathrm{dc}} = 1$ and crosses into the stable limit cycle regime, the performance deteriorates. This behavior is consistent with the results for OP2, where operating in this region led to poor performance. 
Within the stable limit cycle regime, however, a slight improvement is observed for larger values of $I_{\mathrm{dc}}$ and $K_{\mathrm{inj}}$.
Although it is not shown in the current analysis, it is worth
noting that this trend is also reflected in the OP2 case: for $I_{\mathrm{dc}}=1.5$, higher values of $K_{\mathrm{inj}}$ 
yield a marginally better performance, and a similar improvement is observed for larger dc currents (e.g. $I_{\mathrm{dc}} = 3$).
The performance also appears to be very stable if we consult Fig.~\ref{fig:Idc_vs_Kinj}b, where these regions have small standard deviations close to $\pm 0.002$. We can  understood this by considering the dynamics in the $(V_{\mathrm{output}}, \sin\phi_{\mathrm{output}})$ plane, where increased exploration of the phase space is observed. Notably, for dc currents below and close to the infinite-period bifurcation (SNIC) and for small values of $K_{\mathrm{inj}}$, multiple regions yield satisfactory $\mathrm{NRMSE}_{\mathrm{test}}$ values. In these regions, $\mathrm{NRMSE}_{\mathrm{test}}$ is consistently around $0.071$.
Moreover, this performance is highly robust, as indicated in Fig.~\ref{fig:Idc_vs_Kinj}b, where the corresponding standard deviations are small, on the order of $10^{-3}$. Finally, we observe that for $I_{\mathrm{dc}}$ close to the SNIC bifurcation, increasing values of $K_{\mathrm{inj}}$ lead to larger standard deviations and reduced performance of the JJ reservoir.

\subsection{AC input mask}
\label{sec:sinusoidal_mask}
In Sec.~\ref{sec:hyperparam}, we showed that $\theta=1$ provides a suitable
choice for all operating points. In the parameter scan for OP1 (Fig.~\ref{fig:Nodes_theta_scan_col}a), smaller node processing times can also yield satisfactory performance. However, we impose a lower bound of $\theta=1$, as it represents the smallest value compatible with current electronic hardware, in particular due to the rapid switching required for the input mask (see Appendix~\ref{appendix_c}).
It should be noted that even this cutoff corresponds to relatively fast switching. This raises the question of whether an alternative masking strategy could be employed for the JJ reservoir. 

The role of the random mask is to effectively exploit the nonlinearity of the dynamical system and ensure that the high-dimensional phase space is sufficiently explored. One may ask whether, instead of using randomly varying mask values, a structured transformation could be employed. While chaotic masks \cite{KUR18} or trainable masks \cite{BOG25} have already been used, a natural candidate is a sinusoidal function, motivated by the pendulum-like dynamics of the JJ~\cite{STR94}. Such a mask provides a physically consistent harmonic modulation of the input and can be interpreted as introducing an additional ac component to the input current.
Also, in contrast to random masking, this approach provides a smooth and continuous variation across virtual nodes, avoiding abrupt changes at each interval $\theta$.

With the introduction of a sinusoidal mask, Eq.~\eqref{eq:JJ_system_Kinj1} becomes
\begin{equation}
\mathsf{V}' = K_{\mathrm{inj}}\, \mathrm{I}(k)\, \sin(\omega \tau) + I_{\mathrm dc} - \sin\phi - \alpha \mathsf{V},
\label{eq:JJ_AC_Kinj}
\end{equation}
where $\mathrm{I}(k)$ denotes the $k$-th input sample of the time series, and $\omega$ is the angular frequency of the mask.
From Eq.~\eqref{eq:JJ_AC_Kinj}, we observe that, within each virtual node of duration 
$\theta$, the injected current is no longer constant but varies continuously as 
$\sin(\omega \tau)$. The amplitude of this modulation is set by $\mathrm{I}(k)$ and remains constant over one clock cycle $T=N_V\cdot\theta$, while the sinusoidal term introduces a smooth temporal variation across the virtual nodes.

\begin{figure}[tb] 
\begin{overpic}[width=0.8\columnwidth]{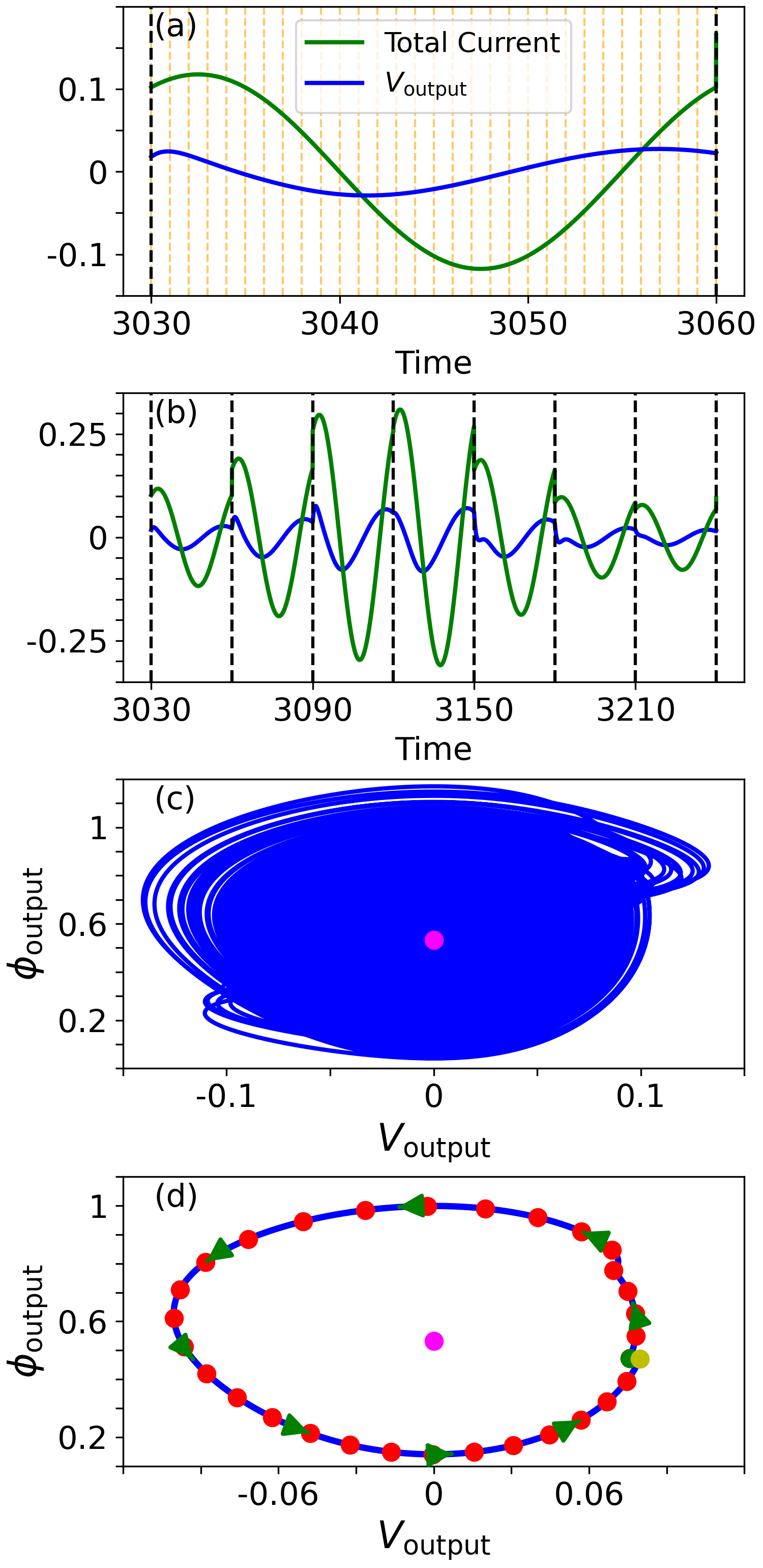}
\end{overpic}
\caption{In (a), the time series of $V_{\mathrm{output}}$ (blue) is shown together with the corresponding virtual nodes (orange dashed lines) and the total input current (green). Panel (b) shows the same quantities over a larger number of clock cycles. In (c), trajectories (blue) in the $(V_{\mathrm{output}},\phi_{\mathrm{output}})$ plane are shown for many clock cycles using the sinusoidal mask. The magenta dot denotes the stable fixed point corresponding to the dc current $I=0.5$. Panel (d) shows trajectories over part of a clock cycle. The green and yellow dots mark the start and end points of the trajectory, respectively. Green arrows indicate the direction of evolution, while red dots denote the instances at which a new total current is injected into the JJ reservoir. Parameters are $N_V=30$, $\theta=1$, $K_{\mathrm{inj}}=0.25$, $\lambda=10^{-13}$, and damping $\alpha=1.5$.}
    \label{fig:AC_MASK}
\end{figure}

We implement the sinusoidal masking scheme using a setup analogous to OP1, where the dc bias current $I_{\mathrm{dc}}=0.5$ places the unperturbed JJ dynamics in the stable fixed point regime. In this case we choose $N_V = 30$, $\theta=1$, $K_{\mathrm{inj}}=0.25$, and $\lambda=10^{-13}$.
The output voltage $V_{\mathrm{output}}$ (blue) is shown together with the mean-subtracted total input current (green) over $7$ clock cycles in Fig.~\ref{fig:AC_MASK}a and b. Owing to the sinusoidal mask, the current exhibits a modulation whose amplitude varies across clock cycles (black dashed lines).

Furthermore, for a single clock cycle shown in Fig.~\ref{fig:AC_MASK}a, the total current is no longer constant within each node interval $\theta$, in contrast to the previous cases (e.g., Fig.~\ref{fig:V_timeseries_V_Phi_plot}a for OP1). We also note that the period of the sinusoidal mask, $2\pi/\omega$, is chosen to match the reservoir clock cycle, $T = N_V\cdot\theta = 30 \cdot 1$. 
A first observation is that the JJ reservoir closely follows the input, suggesting that this configuration can yield good performance.

As in Sec.~\ref{sec:JJ_Response}, we examine the trajectories in the $(V_{\mathrm{output}}, \phi_{\mathrm{output}})$ plane. In Fig.~\ref{fig:AC_MASK}c, the trajectories form a structure that explores a large portion of the phase space, similar to what was observed in Fig.~\ref{fig:V_timeseries_V_Phi_plot}g, and remain centered around the corresponding fixed point. 
Compared to the random masking case, the trajectories obtained with the sinusoidal mask appear noticeably smoother. This is further illustrated in the zoomed-in view for a smaller number of inputs shown in Fig.~\ref{fig:AC_MASK}d. The resulting performance is $\mathrm{NRMSE}_{\mathrm{test}} = 0.079 \pm \textrm{1e-03}$ over 10 different realizations of the Lorenz system. Unlike the random masking case, where performance values are averaged over different mask realizations, the sinusoidal mask corresponds to a single deterministic realization. A more robust assessment would require additional simulations.

This example demonstrates that a sinusoidal mask can be employed in the JJ reservoir, with implications for the lower bound previously imposed on $\theta$ due to electronic limitations. For a sinusoidal mask with a period matched to the clock cycle $T = 30$, Eq.~\eqref{eq:time_dimens} yields $T_{\mathrm{per}} = 6.591\,\mathrm{ps}$ (or $151.72\,\mathrm{GHz}$), which is closer to current hardware capabilities. 
An additional degree of freedom arises from the relation between the mask period and the clock cycle, suggesting further optimization (e.g., using fractions or multiples of $T$). However, it is important to note that sinusoidal driving fundamentally alters the JJ dynamics, leading to more complex behavior~\cite{kostur2008anomalous,gross2016applied,mishra2021neuron} and rendering the previously used stability diagram inapplicable. For instance, AC driving can induce phenomena such as bursting. These effects should be carefully considered in future investigations.  

\section{Conclusions}
\label{sec:conclusions}
We have demonstrated for the first time that a single Josephson junction can serve as a physical substrate for reservoir computing. By examining different dynamical regimes, we showed how the interplay between intrinsic JJ dynamics and input strength determines the reservoir performance. Optimal performance is achieved when the unperturbed JJ operates in a stable fixed-point regime while remaining sufficiently responsive to the input. In contrast, in the stable oscillations regime, the intrinsic dynamics dominate and the influence of the input is weak, leading to reduced computational performance.

Importantly, unlike conventional time-multiplexed reservoirs, our scheme operates without an explicit delay loop. 
Instead, memory arises from the intrinsic dynamics of the JJ, where the system state encodes 
recent inputs and the nonlinear exploration of phase space enriches the reservoir response.
Compared with JTL-based reservoirs~\cite{rowlands2021reservoir,watanabe2024numerical}, our approach relies on a simpler time-multiplexed architecture, avoiding the spatially extended network of coupled junctions while allowing the role of the JJ's intrinsic dynamics in computational performance to be directly assessed.

We also introduced an alternative input masking approach based on a sinusoidal function. In contrast to the commonly used random masks, this AC mask provides a continuous modulation of the input, avoiding rapid switching at each mask interval. This is better aligned to the capabilities of current electronic devices and enables a more physically consistent driving of the JJ, while maintaining competitive performance.

These results demonstrate that Josephson junctions are a promising platform for reservoir computing, offering fast operation and low losses. It is worth noting that the underlying reservoir mechanism is not specific to Josephson junctions. The dynamical analogy to a driven nonlinear pendulum suggests that it can in principle be extended to a broader class of nonlinear oscillators. Josephson junctions nevertheless offer a technologically attractive realization through their ultrafast dynamics, direct electrical control, and established use in superconducting technologies.

Future work could explore the incorporation of delay through input or feedback to enhance memory and performance, as well as the application to additional benchmark tasks~\cite{wringe2025reservoir}. Furthermore, networks of multiple Josephson junctions may provide increased dynamical complexity and improved nonlinear processing capabilities.



\appendix
\section{Benchmark Task and Numerical Setup}
\label{appendix_a}
The Lorenz 63 system is used in this work as a benchmark to evaluate the performance of the JJ reservoir. The Lorenz equations are given by~\cite{Lorenz2004}:
\begin{equation}
	\begin{aligned}
		\dot{X} &= \sigma(Y - X), \\
		\dot{Y} &= rX - Y - XZ, \\
		\dot{Z} &= XY - bZ,
		\label{eq:Lorenz_system}
	\end{aligned}
\end{equation}
where the parameters are chosen as $\sigma = 10$, $b = 8/3$, and $r = 28$, ensuring chaotic dynamics.
Numerical integration of both the Lorenz 63 system and the JJ reservoir was performed using the fourth-order Runge--Kutta method with an integration step equal to $0.01$.

Since the $X$ time series spans the interval $(-19,19)$, i.~e. approximately one order of magnitude larger than the range of the bias current $I_\mathrm{dc}$ of our Josephson Junction (see Fig.~\ref{fig:bif_diagram} in the main text), we rescale the $X$ values by a factor $g_{\text{scale}} = 0.1$ to ensure comparable magnitudes. Consequently, the scaled $X$ values lies within $(-1.9,1.9)$. 

The scaled Lorenz time series $X(t)$ was subsequently sampled with a step size $\Delta t= 0.1$, yielding the discrete sequence $X(k)$, whose length corresponds to the required number of training or testing samples. For the one-step-ahead prediction task $X(t) \rightarrow X(t+\Delta t)$, the input to the reservoir, $I(k)$ in Eq.\ref{eq:I}, is the sequence $X(k)$, while the target output is $X(k+1)$. It should be noted that the sampling interval significantly affects the difficulty of the prediction task~\cite{jaurigue2024reducing}. In addition, buffer intervals are introduced before both the training and testing phases, during which the input is fed into the JJ reservoir in order to initialize the system dynamics.

\section{Parameter Validation}
\label{appendix_b}

\begin{figure*}[t]
\centering
\vspace{1.5em}
\begin{overpic}[width=0.9\textwidth]{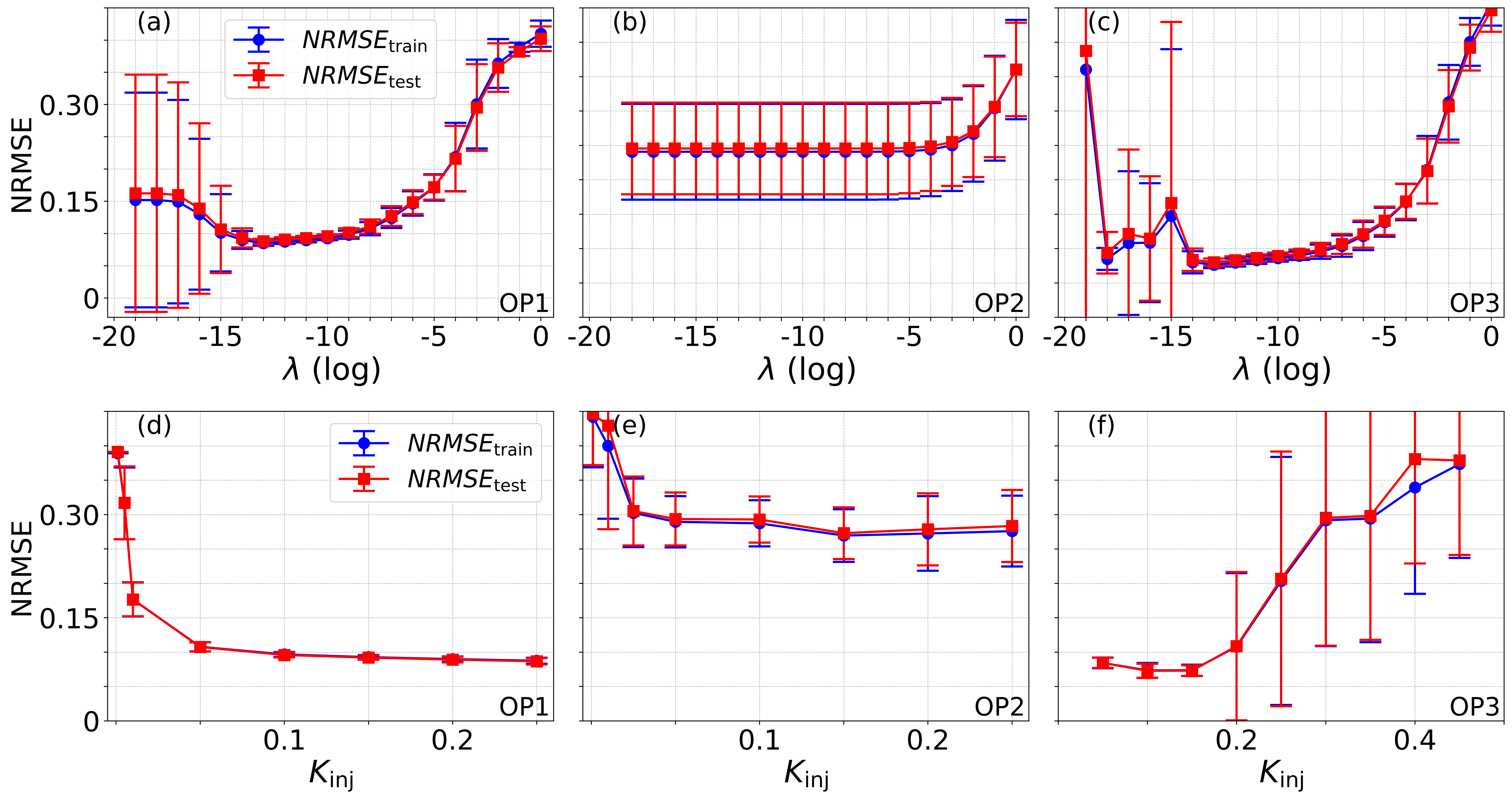}
\end{overpic}
\caption{Mean $\mathrm{NRMSE}{\mathrm{train}}$ (blue) and $\mathrm{NRMSE}{\mathrm{test}}$ (red), together with their standard deviations, computed over $15$ realizations with different random masks, as functions of the ridge parameter $\lambda$ (top row) and the injection strength $K_{\mathrm{inj}}$ (bottom row). OP1 corresponds to Figs.~(a) and (d) with $N_\mathrm{v}=20$ and $\theta=1$. In (a), $K_{\mathrm{inj}}=0.25$ (chosen arbitrarily), while in (d), $\lambda=10^{-13}$. OP2 corresponds to Figs.~(b) and (e) with $N_\mathrm{v}=70$ and $\theta=1$. In (b), $K_{\mathrm{inj}}=0.25$ (chosen arbitrarily), while in (e), $\lambda=10^{-13}$. OP3 corresponds to Figs.~(c) and (f) with $N_\mathrm{v}=20$ and $\theta=1$. In (c), $K_{\mathrm{inj}}=0.1$ (chosen arbitrarily), while in (f), $\lambda=10^{-13}$. In (f) values $K_{\mathrm{inj}}
\leq 0.025$ are not shown, since the SNIC bifurcation cannot be crossed in this range and the resulting dynamics effectively correspond to OP1.}
\label{fig:Lambda_Kinj_scan}
\end{figure*}

In our analysis in Sec.~\ref{sec:hyperparam}, the injection strength was chosen from a dynamical point of view, while the regularization parameter was fixed to $\lambda = 10^{-13}$. 
To verify that these choices are appropriate, we perform parameter scans in both $\lambda$ and $K_{\mathrm{inj}}$, as shown in Fig.~\ref{fig:Lambda_Kinj_scan}. The figure shows the mean $\mathrm{NRMSE}_{\mathrm{train}}$ (blue) and $\mathrm{NRMSE}_{\mathrm{test}}$ (red), along with their standard deviations, computed over $20$ realizations with different random masks. Panels (a)–(c) illustrate the dependence on $\lambda$, while panels (d)–(f) show the effect of $K_{\mathrm{inj}}$.

We observe that, for all operating points, $\lambda = 10^{-13}$ yields consistently low $\mathrm{NRMSE}_{\mathrm{test}}$, supporting our initial choice. 
Since OP2 exhibits generally poor performance due to its limit cycle dynamics, the impact of $\lambda$ here is very weak.
Regarding the input scaling, small values of $K_{\mathrm{inj}}$ lead to suboptimal performance for OP1 and OP2, whereas OP3 shows increased sensitivity at larger $K_{\mathrm{inj}}$. This behavior can be attributed to more frequent transitions into dynamically complex regimes as the input strength increases.

\section{Physical Time Scales}
\label{appendix_c}

Starting from Eq.~\eqref{eq:RCSJ_eq_2}, the dimensionless damping parameter is defined as
\begin{equation}
\alpha=\sqrt{\frac{\hbar}{2e I_c R^2 C}},
\end{equation}
which gives
\begin{equation}
C=\frac{\hbar}{2e I_c R^2 \alpha^2}.
\end{equation}

Using the dimensionless time transformation
\begin{equation}
\tau=\sqrt{\frac{2e I_c}{\hbar C}}\, t,
\end{equation}
and substituting the above expression for \(C\), we obtain
\begin{equation}
\tau=\frac{2e I_c R \alpha}{\hbar}\, t,
\end{equation}
or equivalently
\begin{equation}
t=\frac{\hbar}{2e I_c R \alpha}\,\tau.
\label{eq:normal_time}
\end{equation}

Throughout this work, we use \(\alpha=1.5\). Taking a typical Josephson junction voltage scale of \(I_cR=1\,\mathrm{mV}\),

Eq.~\eqref{eq:normal_time} yields
\begin{equation}
t \approx 0.2197\,\tau\,\mathrm{ps}.
\label{eq:time_dimens}
\end{equation}

Equation~\eqref{eq:time_dimens} shows that one unit of dimensionless time corresponds to approximately \(0.22\,\mathrm{ps}\), reflecting the intrinsic THz operating frequencies of Josephson junctions. While this enables ultrafast reservoir operation, it also poses practical challenges. In particular, implementing random-mask switching at every virtual node would require modulation on picosecond timescales.
By comparison, modern electronic hardware operating around \(100\,\mathrm{GHz}\)~\cite{liou2024gigahertz} corresponds to switching times on the order of \(10\,\mathrm{ps}\). Using Eq.~\eqref{eq:time_dimens}, this translates to \(\theta \approx 45.5\). Such large values of \(\theta\) would substantially increase the required simulation time and make systematic hyperparameter scans computationally demanding.
For this reason, in the present work we restrict the optimization to the range \(1 \leq \theta < 2\). Although these values still correspond to very fast physical timescales, they allow efficient numerical exploration while remaining relevant for future ultrafast hardware implementations.

\end{document}